\documentclass[12pt,showkeys,pra,superscriptaddress,floatfix,nofootinbib,longbibliography]{revtex4-2}

\usepackage{graphicx}
\usepackage{color}
\usepackage{amsfonts}
\usepackage{eurosym}
\usepackage{xspace}

\renewcommand{\d}{{\rm d}}

\newcommand {\E}  {\varepsilon}
\newcommand {\om} {\omega}
\newcommand {\Om} {\Omega}

\newcommand{\MBNExplorer} {\textsc{MBN Explorer}\xspace}

\begin{document}

\title{The impact of ionising collisions on channeling and radiation emission
for high-energy electrons and positrons}


\author{Andrei V. Korol}
\email[]{korol@mbnresearch.com}
\affiliation{MBN Research Center, Altenh\"{o}ferallee 3, 60438 Frankfurt am Main, Germany}

\author{Andrey V. Solov'yov}
\email[]{solovyov@mbnresearch.com}
\affiliation{MBN Research Center, Altenh\"{o}ferallee 3, 60438 Frankfurt am Main, Germany}

\begin{abstract}
This paper presents a quantitative analysis of the impact
of inelastic collisions with atoms in a crystalline environment
on the channeling efficiency and intensity of the channeling radiation
for high-energy electrons and positrons passing through oriented
crystalline targets.
This analysis is based on numerical simulations of the channeling
process,
which were performed using the MBN Explorer software package.
Ionising collisions are considered random, fast and local events, and are
incorporated into the classical relativistic molecular dynamics framework
according to the previously described algorithm.
The case studies presented refer to 10 GeV electrons and positrons
incident on single crystals of diamond and silicon, oriented along
the (110) and (111)
planes, with thicknesses of up to 1 mm for electrons and 6 mm for
positrons.%
To elucidate the role of ionising collisions, simulations were performed
with and without accounting for them.
It is shown that, for electrons, both approaches lead to similar results
with regard to both the channelling efficiency and the radiation
intensity.
In practical terms, this means that numerical simulations can be
carried out without accounting for ionising collisions, which are much
faster yet produce similar results.
For positrons, the ionising collisions reduce significantly the
channeling efficiency.
However, their impact on the radiation intensity strongly depends on
the opening angle of the cone within which the radiation emission
is collected.
A quantitative analysis of this feature is presented in the paper.

\end{abstract}


\maketitle

\section{Introduction  \label{Introduction}}

The dynamics of high-energy charged particles propagating through
crystalline targets, as well as the radiation emitted by the particles,
are sensitive to the orientation of the incoming beam with respect to
the main crystallographic directions of the target.
Projectiles incident at small angles to the crystal planes (or axes)
experience channelling motion along the planar (or axial) direction,
due to the collective action of the electrostatic fields of the
lattice atoms \cite{Lindhard}.
The study of the channeling of ultra-relativistic particles in oriented
crystals is a well-established field of research
\cite{Uggerhoj:RPM_v77_p1131_2005,BiryukovChesnokovKotovBook,%
ChannelingBook2014}.

A potential application of the channeling motion of ultra-relativistic
electrons and positrons in crystals of different shapes and
orientations is the creation of
novel types of intense gamma-ray light sources (LS) operating in the
photon energy range from hundreds of keV to GeV \cite{CLS-book_2022},
an area of current interest in many fields including fundamental science,
industry, biology and medicine
\cite{AlbertThomas_PlasmaPhysContrFusion_v58_103001_2016,%
Rehman_EtAl_ANE_v105_p150_2017,Kraemer_EtAl-ScieRep_v8_p139_2018,
NextGenerationGammaRayLS2022,BudkerEtAl:AnnPhys_v534_2100284_2022}.
As shown in Ref. \cite{KorolSushkoSolovyov:PRAB_v27_p100703_2024},
the intensity of radiation from crystal-based LS (CLS)
can exceed that achievable at modern laser-Compton facilities.

The goal of the current Horizon Europe EIC-Pathfinder-2021 project,
TECHNO-CLS \cite{TECHNO-CLS}, is the practical realisation of CLSs.
This challenging task combines the development of technologies for preparing
crystal samples with an experimental programme for designing and
manipulating particle beams, as well as characterising emitted
radiation.
Another key element is theoretical analysis and advanced computational
modelling of the processes involved.

A quantity that is important for the characterisation of a CLS
is the spectral distribution of radiation emitted by incident particles
as they propagate through the crystal in channeling mode.
The elastic and inelastic collisions\footnote{Below in the paper,
inelastic collisions that lead to the excitation or ionisation of atoms
will be referred to as 'ionising collisions'.}
of the particles with the constituent atoms
lead to a gradual decrease in the number of channeling particles,
consequently decreasing the radiation intensity.
In a recent paper \cite{SushkoKorolSolovyov:NIMB_v569_165911_2025},
a numerical analysis was presented of the impact of ionising collisions
on the channelling efficiency and  photon emission for
270 – 1500 MeV electrons and 530 MeV positrons\footnote{Such beams
are available at the MAinz MIkrotron (MAMI) facility
\cite{Jankowiak:EPJA_v28_p149_2006,Backe:NIMA_v1059_168998_2024,
BackeEtAl:JINST_v13_C04022_2018}.}
channeling in single crystals of diamond, silicon and germanium.
The role of ionising collisions was elucidated by performing simulations
with and without accounting for them.
The results obtained demonstrate that both approaches yield highly similar
results for electrons.
For positrons, however, it was observed that a significant reduction in
channelling efficiency due to ionising collisions does not result in a
corresponding change in the spectral distribution of the channelling
radiation emitted within a wide cone along the incident beam.
While a qualitative explanation of this phenomenon was provided, it
was not supported by quantitative analysis.

 In the present paper, the methodology developed in
 Ref. \cite{SushkoKorolSolovyov:NIMB_v569_165911_2025} is applied to carry
 out an accurate numerical analysis of the influence of ionising
 collisions on electrons and positrons with much higher energies
 (10 GeV; the SLAC facility
 \cite{Yakimenko-EtAl:PR-AB_v22_101301_2019,%
 FACETII_Technical_Design_Rep-2016})
channelled  through thick diamond and silicon crystals (up
to 1 and 6 millimetres for  electrons and positrons, respectively)
oriented along the (110) and (111)
 planar directions.
A more systematic study than that performed in the cited paper is carried
out to investigate the sensitivity of changes in radiation emission
spectra
to ionising collisions, with respect to:
(i) the charge of the projectile particle;
(ii) the type and thickness of the crystalline target;
and (iii) the angle of the emission cone.
Additionally, for positrons the results of a quantitative analysis are
presented that explain why the decrease in the intensity of
the channeling radiation can be much less pronounced than that in
the channeling efficiency.

Numerical simulations of the trajectories of ultra-relativistic
electrons and positrons in a crystalline medium were performed
with an atomistic level of accuracy
within the relativistic classical molecular dynamics (Rel-MD)
framework
\cite{MBN_ChannelingPaper_2013,KorolSushkoSolovyov:EPJD_v75_p107_2021},
using the multi-purpose computer package \textsc{MBN Explorer}
\cite{MBNExplorer_2012}.
 The implemented algorithms allow for the modelling of the passage
 of particles over macroscopic distances with atomistic accuracy,
 accounting for the interaction of a projectile with all the atoms in the
 environment.
 This goes beyond the continuous potential model  \cite{Lindhard}.
 Particle trajectories are generated by accounting for randomness in the
 sampling of incoming projectiles, as well as in the displacement of
 lattice atoms due to thermal vibrations.
Consequently, each trajectory corresponds to a unique crystalline
environment.
Another phenomenon affecting the dynamics of a projectile particle and
contributing to the statistical independence of the simulated trajectories
is ionising collisions, which lead to a random change in the particle’s
velocity direction.
These quantum events are random and occur on the atomic scale in terms of
time and space; therefore, they are incorporated into classical Rel-MD
in accordance with their probabilities.
The methodology implemented in the code to account for the ionising
collisions is detailed in Ref.
\cite{SushkoKorolSolovyov:NIMB_v569_165911_2025}.
The simulated trajectories are used to analyse the channeling efficiency
as well as to characterise the emitted radiation (by calculating the
spectral-angular and spectral distribution of radiation) within the
framework of quasi-classical formalism \cite{Baier}.

\section{Results and discussion  \label{CaseStudies}}

Simulations were performed for positron and electron beams
with an energy of 10 GeV, incident on oriented diamond and
silicon single crystals along the (110) and (111) planes.
The thickness of the crystals ($L$) along the direction of the incident
beam ($z$ direction) was varied from 1 to 6 mm for positrons and
from 0.1 to 1 mm for electrons.
The $y$-axis was aligned with either $\langle 110\rangle$ or
$\langle 111\rangle$ axes depending on the choice of the planar
orientation.
Depending on the choice of planar orientation, the $y$-axis was aligned
with either the $\langle 110\rangle$ or $\langle 111\rangle$ axes.

The divergences of $\phi_{x,y}=10,\, 30$ $\mu$rad used in the
simulations for both electron and positron beams correspond to the
FACET-II beams
\cite{Yakimenko-EtAl:PR-AB_v22_101301_2019,FACETII_Technical_Design_Rep-2016}.
The value of $\phi_y$ is more than twice as small as Lindhard's critical
angle.
Therefore, a sufficiently large fraction of the beam particles is accepted
in the channelling mode at the crystal entrance.
The divergence along the $x$-direction is much smaller than
the natural emission angle $\gamma^{-1} \approx 50$ $\mu$rad.

At the crystal entrance, the initial velocities $v_{x0}$ and $v_{y0}$ of
an incident particle were generated using normal
distributions with the standard deviations $\phi_{x}$ and $\phi_{y}$,
respectively.
The initial coordinates $x_0$ and $y_0$ were generated using uniform
distributions.
The values of $x_0,y_0,v_{x0}$, and $v_{y0}$ served as the initial
conditions when integrating the classical relativistic equations of motion
to calculate the trajectories of particles in an electrostatic field
created by all the atoms in a crystalline medium.
During the passage of the particles, the crystalline environment was
generated dynamically accounting for the random displacement of the atoms
from their nodal positions due to thermal vibrations.

Classical equations of motion do not account for the inelastic scattering
of a projectile particle from individual atoms, which leads to the
excitation or ionisation of the atom.
Since they are fast and local, they can be incorporated into the classical
mechanics framework according to their probabilities.
This approach has already been implemented in \MBNExplorer within the
framework of irradiation-driven molecular dynamics
\cite{Sushko_IS_AS_FEBID_2016}.
A similar methodology has been applied to account for ionising collisions.

At each step of the integration of the equations of motion the probability
of  the projectile's energy loss in an inelastic collision with
quasi-free atomic electrons is calculated.
If the collision occurs, the scattering angles measured with respect to
the instant velocity are calculated.
These values are then used to modify the projectile's velocity at the
start of the next integration step.
The corresponding algoritms are described in
Ref. \cite{SushkoKorolSolovyov:NIMB_v569_165911_2025}.

\subsection{Case study I: 10 GeV electrons  \label{CaseStudy_I}}

Both elastic and inelastic collisions can lead to an increase in the
transverse energy of a channelling particle.
Consequently, the number of particles accepted in the channelling regime
at the
entrance, decreases with the penetration distance $z$.
This dependence can be characterised by considering the ratio
$f_{\rm ch,0}(z) = N_{\rm ch,0}(z)/N_0$, where
$N_0$ stands for the total number of incident particles and
$N_{\rm ch,0}(z)$ is the number of the accepted particles
that channel up to $z$.
The value of $f_{\rm ch,0}(0)$ determines the acceptance rate, representing
the fraction of the particles accepted at the entrance out of the total number
of particles.
A rechannelling event, which is the opposite of dechannelling, can also
occur to the collisions.
In this case, a particle that experiences unrestricted over-barrier motion
loses its transverse energy in the collision.
As a result, it can be captured in the channelling mode at some
point within a crystal.
With account for the rechannelling the number of particles
$N_{\rm ch} (z)$, which move in the channelling mode at distance $z$
is greater than $N_{\rm ch,0}(z)$.

\begin{figure} [h]
\centering
\includegraphics[clip,width=\textwidth]{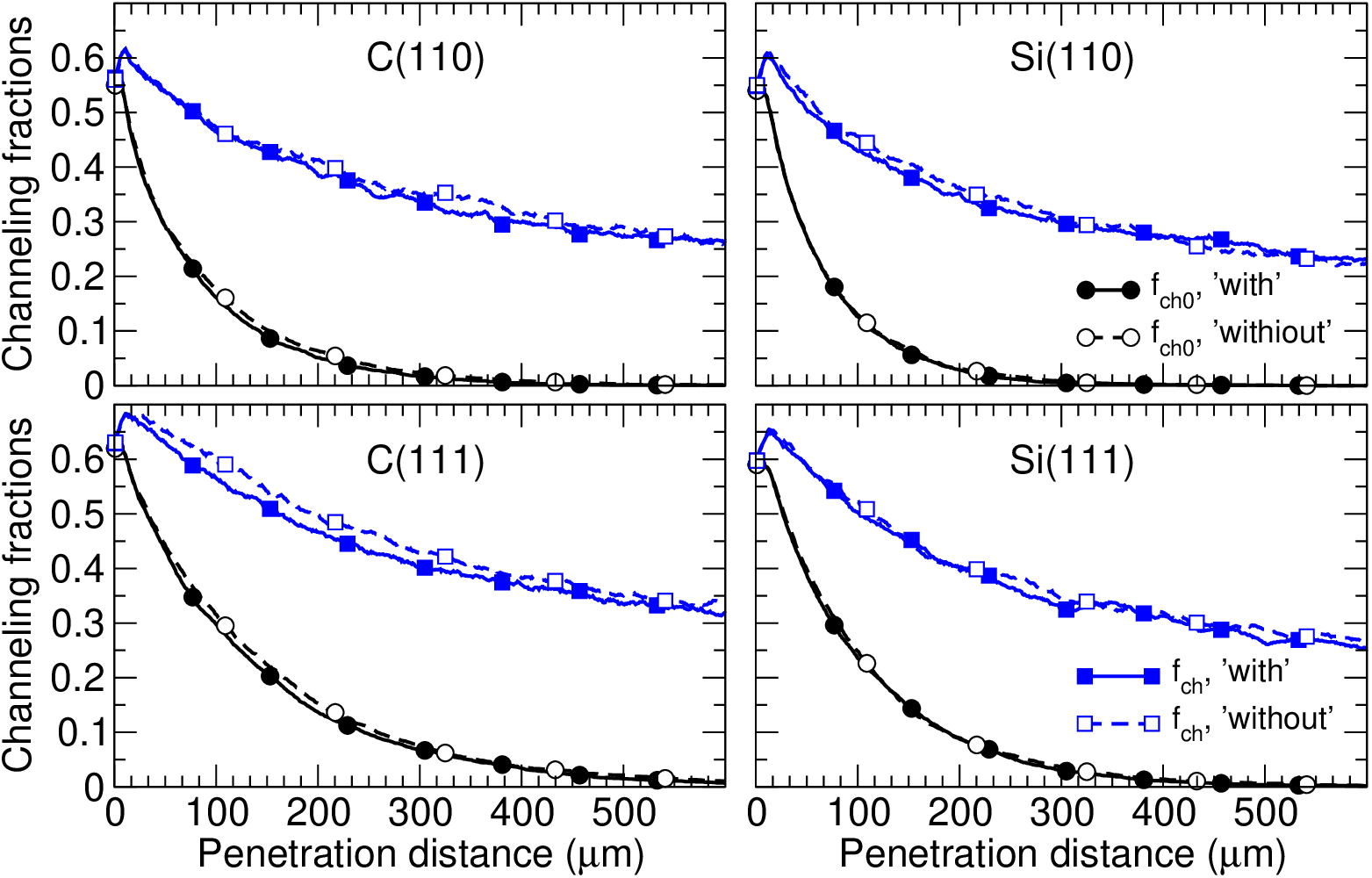}
\caption{
Channeling fractions $f_{{\rm ch}, 0}$ (solid lines) and
$f_{\rm ch}$ (dashed lines) versus penetration distance $z$
for 10 GeV electrons in oriented diamond (left column) and silicon
(right column) crystals.
The upper row corresponds to the crystal orientation along
the (110) plane, the lower row - along the (111) plane.
Curves with filled symbols (circles and squares) show the results obtained
with account for ionising collisions.
Curves with the open symbols represent the dependencies calculated
without ionising collisions (labelled 'without').
}
\label{Figure01.fig}
\end{figure}

Figure \ref{Figure01.fig} shows the dependence of the channelling
fractions $f_{\rm ch,0}(z)$ (solid curves) and $f_{\rm ch}(z)$
(dashed curves) for 10 GeV electrons.
In each graph, the fractions calculated with ionising collisions
accounted for (marked ’with’ in the common legends) are
compared with those calculated without the ionising collisions.
The main feature seen in all graphs is that the ionising collisions
do not lead to a significantly alter either fraction.
This is consistent with an earlier observation for much lower energy
electrons \cite{SushkoKorolSolovyov:NIMB_v569_165911_2025} that the
ionising collisions can be disregarded in the quantitative description
of the dechannelling and rechannelling processes.
The cited paper explains this phenomenon by stating that channelling
motion of negatively charged particles occurs in the vicinity of the
atomic planes.
Consequently, collisions with crystal atoms occur with comparatively small
impact parameters for which the change in the projectile particle's
transverse momentum  is mainly due to elastic scattering from the static
atomic potential, rather than inelastic scattering from the atomic
electrons.

The probability of an inelastic collision is proportional to the
electron density, $n$.
Both diamond and silicon crystals have a cubic lattice of the diamond
group with a unit cell containing eight atoms.
Therefore, if one estimates the electron density as $8Z/a^3$
(where $Z$ is the nuclear charge and $a$ is the lattice constant,
one finds that the density in the diamond crystal ($Z=6$, $a=3.56$ \AA)
is approximately 1.5 times higher than in the silicon crystal
($Z=14$, $a=5.43$ \AA).
This consideration explain why the
impact of inelastic collisions on the channelling fractions
is more noticeable for the diamond target.

The spectral distribution of radiation emitted within the cone along the
incident beam with an opening angle of $\theta_0$ was calculated
numerically for each simulated trajectory, following the algorithm
outlined in Ref. \cite{MBN_ChannelingPaper_2013}.
The resulting spectrum was obtained by averaging the individual spectra
over all trajectories.

\begin{figure} [h]
\centering
\includegraphics[clip,width=\textwidth]{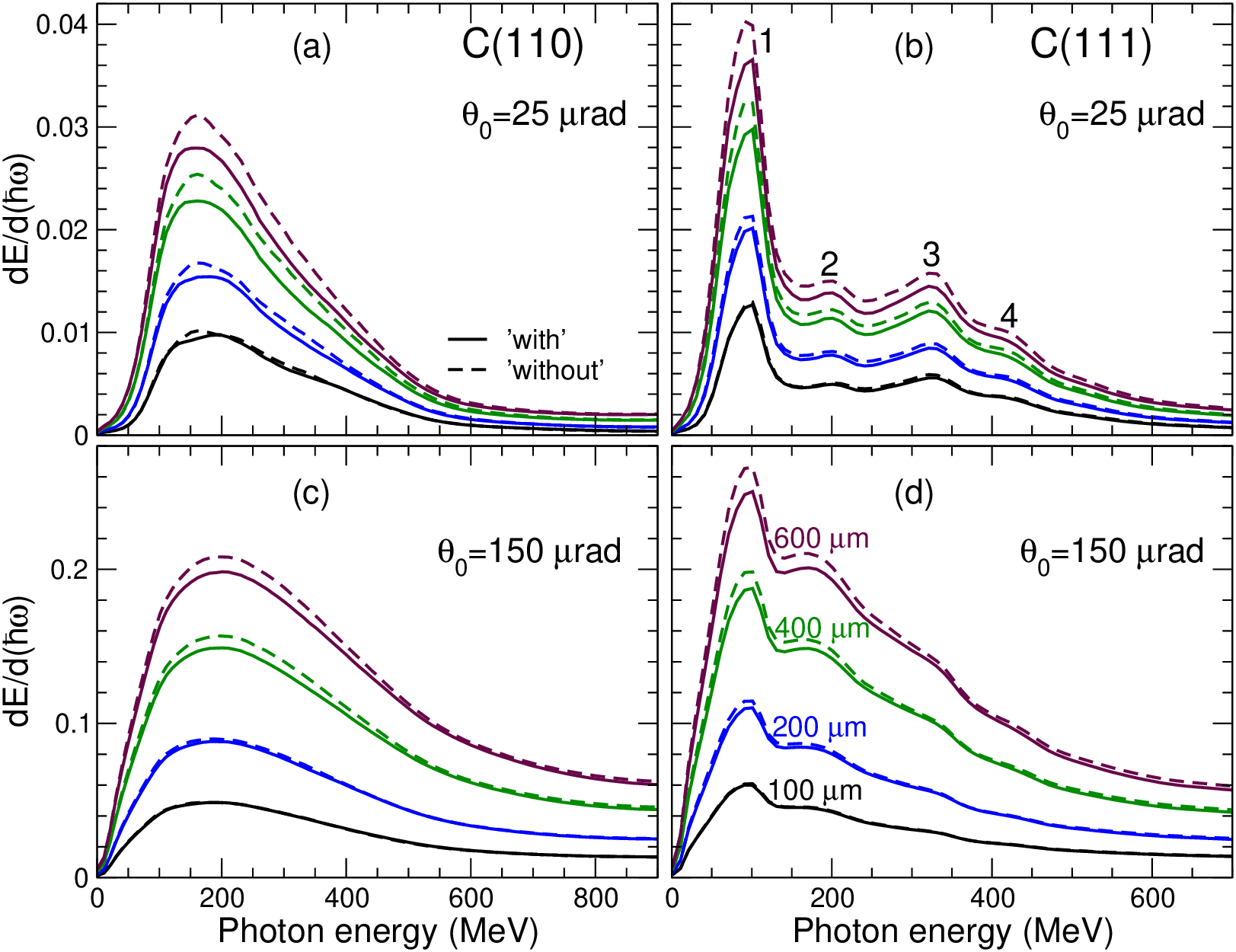}
\caption{
 Spectral distribution of radiation emitted by 10 GeV electrons
channeled in the diamond crystal oriented along the (110) and (111) planes
(left and right columns, respectively).
The upper and lower rows correspond to the emission cones
$\theta_0=25$ and 150 $\mu$rad, respectively.
The solid and dashed curves show the results obtained with and without
the ionising collisions being accounted for.
The spectra shown refer to the crystal thicknesses of 100, 200,
400 and 600 microns, as indicated in the right-bottom graph.
In graph (b), labels 1-4 mark the features discussed in the text.
}
\label{Figure02.fig}
\end{figure}
\begin{figure} [h]
\centering
\includegraphics[clip,width=\textwidth]{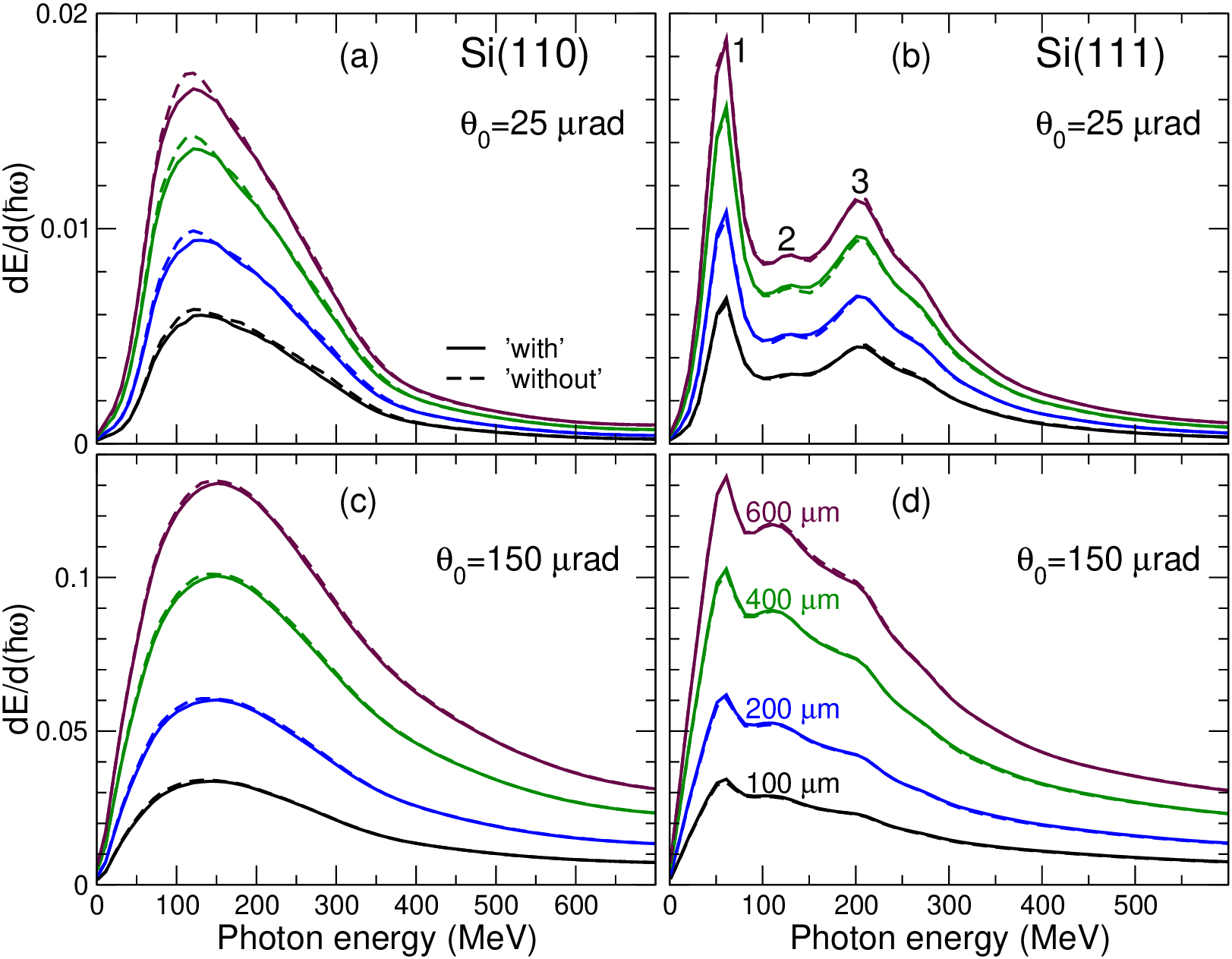}
\caption{
Same as in Fig. \ref{Figure02.fig} but for oriented silicon crystal.
}
\label{Figure03.fig}
\end{figure}

Figs. \ref{Figure02.fig}  and \ref{Figure03.fig} show the spectral
distributions for electrons in the diamond and silicon targets,
respectively.
The spectra are shown within the photon energy range in which
the channelling radiation is the dominant mechanism.
In both figures the left and right columns correspond to the (110)
and (111) crystal orientations, respectively.
The graphs in the top row refer to the narrow emission cone with an angle
$\theta_0=25$ $\mu$rad, which is approximately half the size of
the natural emission cone $\gamma^{-1}$.
The spectra at the bottom were calculated for a much larger cone of
$150$ $\mu$rad, which collects nearly all radiation emitted by the
particles.
In each graph, the four sets of solid and dashed curves correspond to
different crystal thicknesses, as  indicated in graphs (d).

For the purposes of this paper, it is important to note that, as with the
channelling fractions, ionising collisions do not significantly affect the
spectral distributions.
The most significant discrepancy (approximately 10 per cent in the
maximum of the spectra) between the 'with' and 'without' curves occurs
for the diamond targets.
A similar discrepancy was previously observed for electrons with
lower energies (270–1500 MeV)
\cite{SushkoKorolSolovyov:NIMB_v569_165911_2025}.

Another feature evident in Figs. \ref{Figure02.fig}
and \ref{Figure03.fig}
is the difference in the spectral distribution profiles of channeling
radiation emitted by electrons in the (110) and (111) channels.
In the former case, the distribution calculated for each emission cone
has a single, powerful broad maximum and
decreases smoothly away from it.
In the (111) spectra for the smaller emission cone (graphs (b)),
two other maxima
can be identified beyond the first, comparatively narrow maximum,
which is located at approximately
$\hbar\om_{1}=100$ and 60 MeV for the diamond and silicon crystals,
respectively.
These maxima are labelled "1", "2" and "3" in the graphs (b).
The corresponding photon energies, $\hbar\om_{2,3}$, are
approximately two and three times larger than $\hbar\om_{1}$.
The peak intensity of the third maximum is higher than that of
the second maximum.
For the diamond target, the fourth irregularity (labelled "4"
in Fig. \ref{Figure02.fig}(b)) at
$\hbar\om_{4}\approx 4\hbar\om_{1}$ can be identified.
An increase in the emission cone broadens the first and second maxima, and
reduces the third maximum to a hump on the right shoulder of the spectrum
(see graphs (d)).
The peak value at $\hbar\om_{2}$ exceeds those at $\hbar\om_{3}$.

Therefore, Figs. \ref{Figure02.fig} and \ref{Figure03.fig} indicate
that the spectral distributions of radiation emitted by electrons in
the (111)-oriented crystals resemble the features of planar undulator
radiation (see, e.g., \cite{AlferovBashmakovCherenkov1989}).
In particular, these features include:  (i) enhancement of radiation at
equally spaced frequencies (harmonics), which are integer multiples of the
fundamental harmonic; and
(ii) suppression of radiation into even harmonics at small emission
angles.
For (110)-oriented targets these features are smeared out.
A quantitative assessment of these observations, based on statistical
analysis of simulated trajectories, is presented in Appendix
\ref{Appendix01}.

\subsection{Case study II: 10 GeV positrons  \label{CaseStudy_II}}

\begin{figure} [h]
\centering
\includegraphics[clip,width=\textwidth]{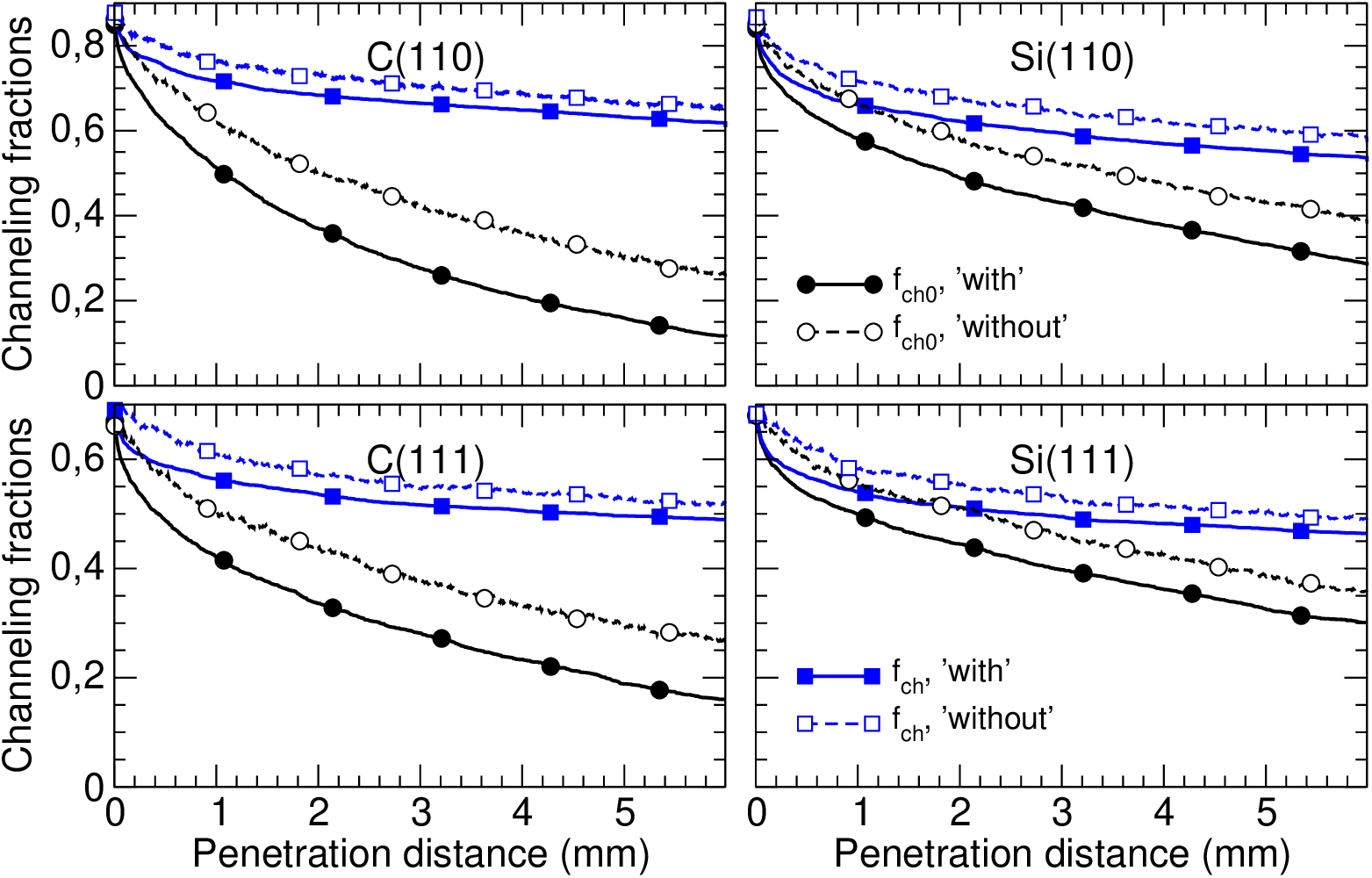}
\caption{
Same as in Fig. \ref{Figure01.fig} but for 10 GeV positrons.
}
\label{Figure04.fig}
\end{figure}

Unlike electrons, positrons predominantly exhibit channelling motion
in the regions between atomic planes.
Consequently, they experience more distant collision with crystal
atoms.
At the distances larger than the average atomic radius,
inelastic channels dominate elastic ones.
It is therefore reasonable to expect that accounting for ionising
collisions will result in a much steeper decrease in the positron
channelling fractions with penetration distance.
This expectation is supported by Fig. \ref{Figure04.fig}, which
presents the channelling fractions calculated for
10 GeV positrons in macroscopically large (up to 6 mm)
oriented diamond and silicon crystals.
The curves shown for the (111) orientation of refer to the 'wide'
part of the  channel, which has a much higher channelling efficiency than
its 'narrow' part (see, for example, Appendix A in
\cite{SushkoKorolSolovyov:NIMB_v569_165911_2025} for illustrative data).

To be noted that the rate at which the fractions decay changes with
respect to the nuclear charge $Z$ when ionising collisions are taken into
account.
If only elastic channels are considered, then the fractions decay faster
in a heavier crystal where elastic cross section (even for distant
collisions) is enhanced due to larger $Z$ value.
As mentioned above, the probability of an inelastic collision is
proportional to the electron density, which is higher in the diamond
crystal.
As a result, the decay rate of the channelling fractions that account for
ionising collisions becomes higher for the lighter crystal.

\begin{figure} [h]
\centering
\includegraphics[clip,width=\textwidth]{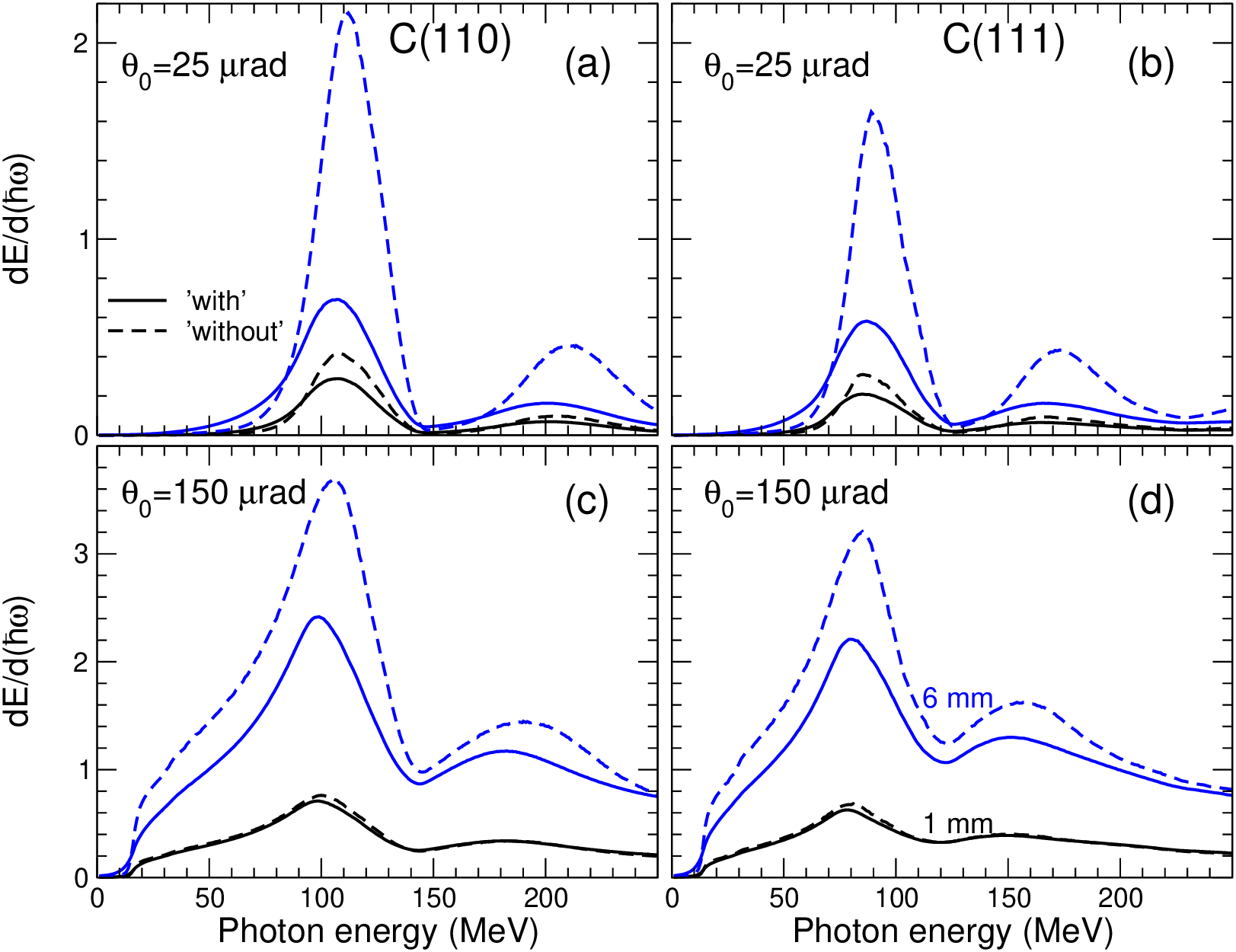}
\caption{
 Spectral distribution of radiation emitted by 10 GeV positrons
channeled in the diamond crystal oriented along the (110) and (111) planes
(left and right columns, respectively).
The upper and lower rows correspond to the emission cones
$\theta_0=25$ and 150 $\mu$rad, respectively.
The solid and dashed curves show the results obtained with and without
the ionising collisions being accounted for.
The spectra shown refer to the crystal thicknesses of 1 and 6
millimeters, as indicated in the right-bottom graph.
}
\label{Figure05.fig}
\end{figure}
\begin{figure} [h]
\centering
\includegraphics[clip,width=\textwidth]{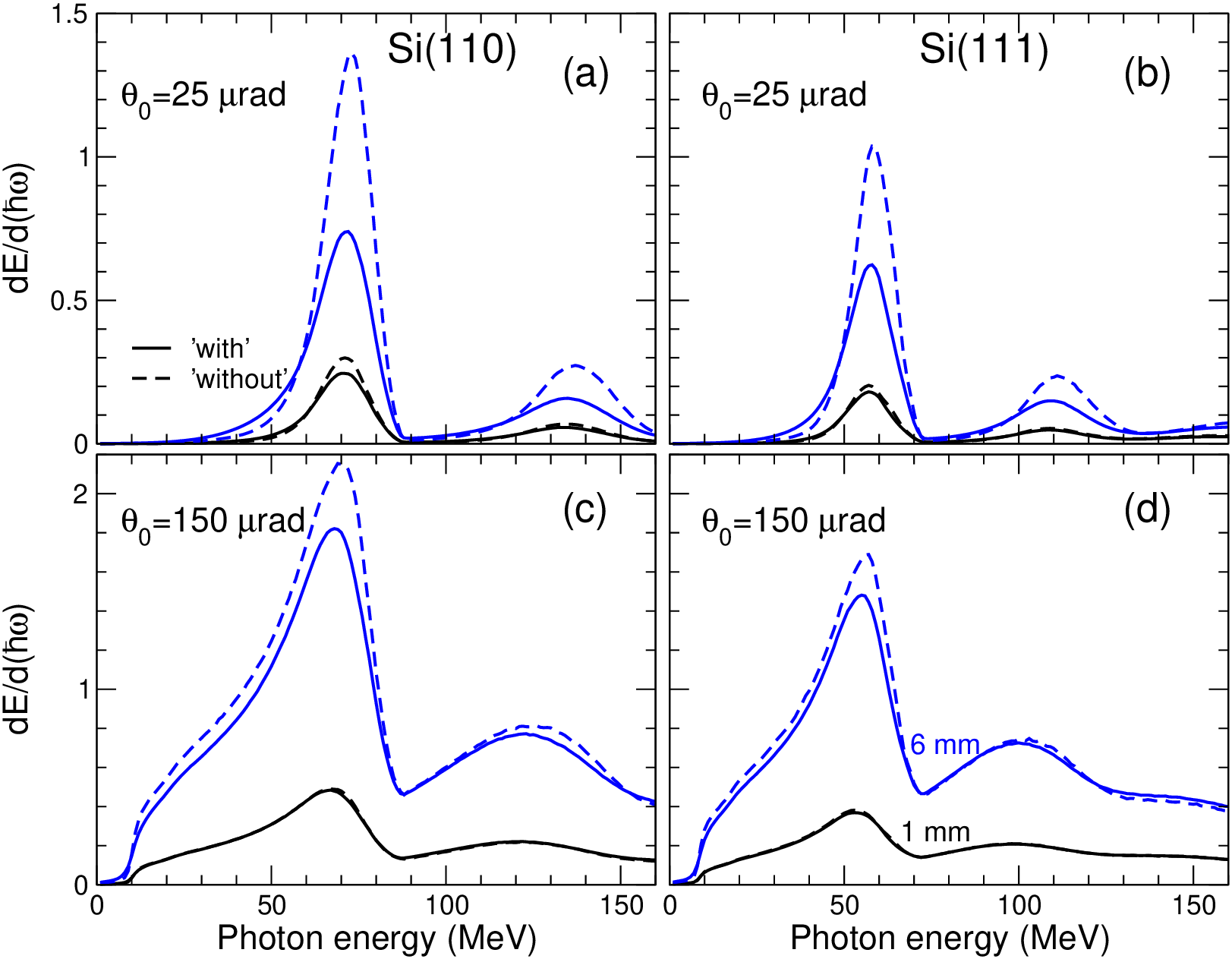}
\caption{
Same as in Fig. \ref{Figure05.fig} but for oriented silicon crystal.
}
\label{Figure06.fig}
\end{figure}

However, the noticeable decrease in the decay rate of the channelling
fractions due to  ionising collisions does not always result in an
equivalent decrease in the intensity of the channelling radiation.
The spectral distributions of the radiation are presented in
Figs. \ref{Figure05.fig} and \ref{Figure06.fig} for the diamond and
silicon targets, respectively.
As with the electron spectra, the left and right columns of each figure
refer to the (110) and (111) crystal orientations, and the top and bottom
rows correspond to spectra calculated for narrow and wide emission cones,
$\theta_0=25$ and 150 $\mu$rad, respectively.

For positrons, the channelling oscillations between neighbouring planes
 are almost harmonic.
Therefore, the spectral distribution of the channelling oscillations
closely resembles that of undulator radiation, where enhancement of the
radiation intensity occurs in the vicinity of equally spaced harmonics
(see, for example,  \cite{Baier}).
For the diamond crystal, the peaks correspond to the first harmonics,
whereas for the silicon crystal, the first two harmonics are clearly
visible within the considered photon energy range.
A detailed analysis of the calculated spectra has shown that,
in all cases,
the main contribution (over 90 per cent) to the first
harmonic peak intensity comes from the fraction of the accepted particles.

When comparing the dependencies calculated with (solid lines) and without
(dashed lines) accounting for ionising collisions, the following features
can be observed:
\begin{enumerate}
\item[(i)]
For the same emission cone value, the difference is more pronounced
for the diamond crystal.

\item[(ii)]
For each crystal and emission cone, the difference increases with
the crystal thickness.

\item[(iii)]
For each crystal and for each thickness,
the difference decreases as the emission cone increases.

\end{enumerate}

To characterize these features quantitatively and for further reference
it Table \ref{Table.01} we list several ratios of the quantities
calculated without and with accounting for ionising collisions.
The ratio of the channelling fractions $f_{\rm ch,0}(z)$ (solid lines in
Fig. \ref{Figure04.fig}) calculated at $z=1$ and 5 mm is notated as
$\xi_{\rm ch, 0}$.
The ratios of the first harmonic peak intensities calculated at
$\theta_0=25$ and 150 $\mu$rad and for the two values of the crystal
thickness ($L=1$ and 6 mm) are notated as $\xi_{25}$ and $\xi_{150}$.

\begin{table}[h]
\caption{Ratios of the quantities calculated
for the positron channelling without and with accounting
for ionising collisions:
(i) $\xi_{\rm ch, 0}$ is the ratio of the channelling fractions
$f_{\rm ch, 0}$ of the accepted particles,
(ii) $\xi_{25}$ and $\xi_{150}$ are the ratios of the first
harmonic peak intensities for the emission cones
$25$ and 150 $\mu$rad, respectively.
Two sets of the ratios correspond to two values of the crystal thickness,
$L$.
}
\begin{tabular}{p{1.5cm}p{1.0cm}p{1.0cm}p{1.0cm}p{0.5cm}p{1.0cm}p{1.0cm}p{1.0cm}}
       &\multicolumn{3}{c}{$L=1$ mm} & &\multicolumn{3}{c}{$L=6$ mm} \\
       &$\xi_{\rm ch, 0}$&$\xi_{25}$&$\xi_{150}$& &$\xi_{\rm ch, 0}$&$\xi_{25}$&$\xi_{150}$ \\
\hline
C(110) & 1.38  &  1.44    &  1.07     & &  5.00 &  3.20    & 1.52
 \\
\hline
C(111) & 1.36  &  1.39    &  1.09     & &  3.02 &  2.83    & 1.45
 \\
\hline
Si(110)& 1.13  &  1.22    &  1.01     & &  1.85 & 1.84     &  1.18
 \\
\hline
Si(111)& 1.08  &  1.12    &  1.04     & &  1.52 & 1.62     & 1.14
\\
\hline
\end{tabular}
\label{Table.01}
\end{table}

Features (i) and (ii), mentioned above, are the direct consequence of
the channelling fractions' dependence on crystal type and penetration
distance.
The intensity of the channelling radiation depends on the number of
channelling particles, which decreases monotonically with distance.
Ionising collisions result in a higher dechannelling rate and thus
decrease the intensity.
As discussed above, the impact of ionising collisions on the dechannelling
rate is greater for the diamond crystal.

The impact of the ionising collisions on the dependence of the
the spectral distribution on the emission cone ( feature (iii) ) is less
obvious.
The data in Table \ref{Table.01} show that, for a smaller emission
cone the values of $\xi_{25}$ are close to the ratio of the channelling
fractions.
However, for the lager value of $\theta_0$ the ratio of the intensities
$\xi_{150}$ is smaller than $\xi_{\rm ch, 0}$ up to a factor greater than
two in the case of a diamond crystal of the larger thickness.
The observation that an increase in the dechannelling rate does not
necessarily lead to a comparable change in the spectral distribution of
radiation emitted in a wide cone along the incoming beam
was also made in \cite{SushkoKorolSolovyov:NIMB_v569_165911_2025}
for 530 MeV positrons.
The qualitative explanation presented there was based on the assumption
that ionising collisions reduce the length of the channelling segment of a
particle's trajectory, while also increasing the amplitude of channelling
oscillations.
The former reduces the intensity of the channelling radiation, while the
latter increases it, resulting in a spectrum that is a balance of these
opposing tendencies.

The quantitative analysis presented below supports
the qualitative explanation.

To carry out this analysis, we make use of the above observations that
the peak of the channelling radiation is mainly due to the radiation
emitted by the particles accepted in the channelling mode of motion at
the entrance of the crystal.
The spectral distribution of the energy radiated by such a particle
and \textit{integrated over all emission angles}
is proportional to (i) the number of periods made, which
is linearly related to the penetration distance $L_{\rm p}$
covered by the particle in the channelling mode, and
(ii) the average square of the Fourier component of the particle’s
acceleration,  $\langle \Om_{\rm ch}^4a_{\rm ch}^2 \rangle$
where $\Om_{\rm ch}$ and $a_{\rm ch}$
are the frequency and the amplitude of the channelling oscillations.
Assuming the harmonic nature of positron channelling
oscillations for a positron, $\Om_{\rm ch}$ becomes
independent of amplitude and is therefore the same for all channelling
particles.
The following proportionality can then be written for
the spectral distribution
$\d E_{\rm acc}/\d (\hbar \om) \equiv I_{\rm acc}$
of radiation emitted by all accepted particles:
\begin{eqnarray}
I_{\rm acc}
\propto
\sum_{j} \left(L_{\rm p}\left\langle a_{\rm ch}^2 \right\rangle\right)_j
\label{CaseStudy_II:eq.01}
\end{eqnarray}
where the sum is carried out over the accepted particles.

\begin{figure} [h]
\centering
\includegraphics[clip,width=0.9\textwidth]{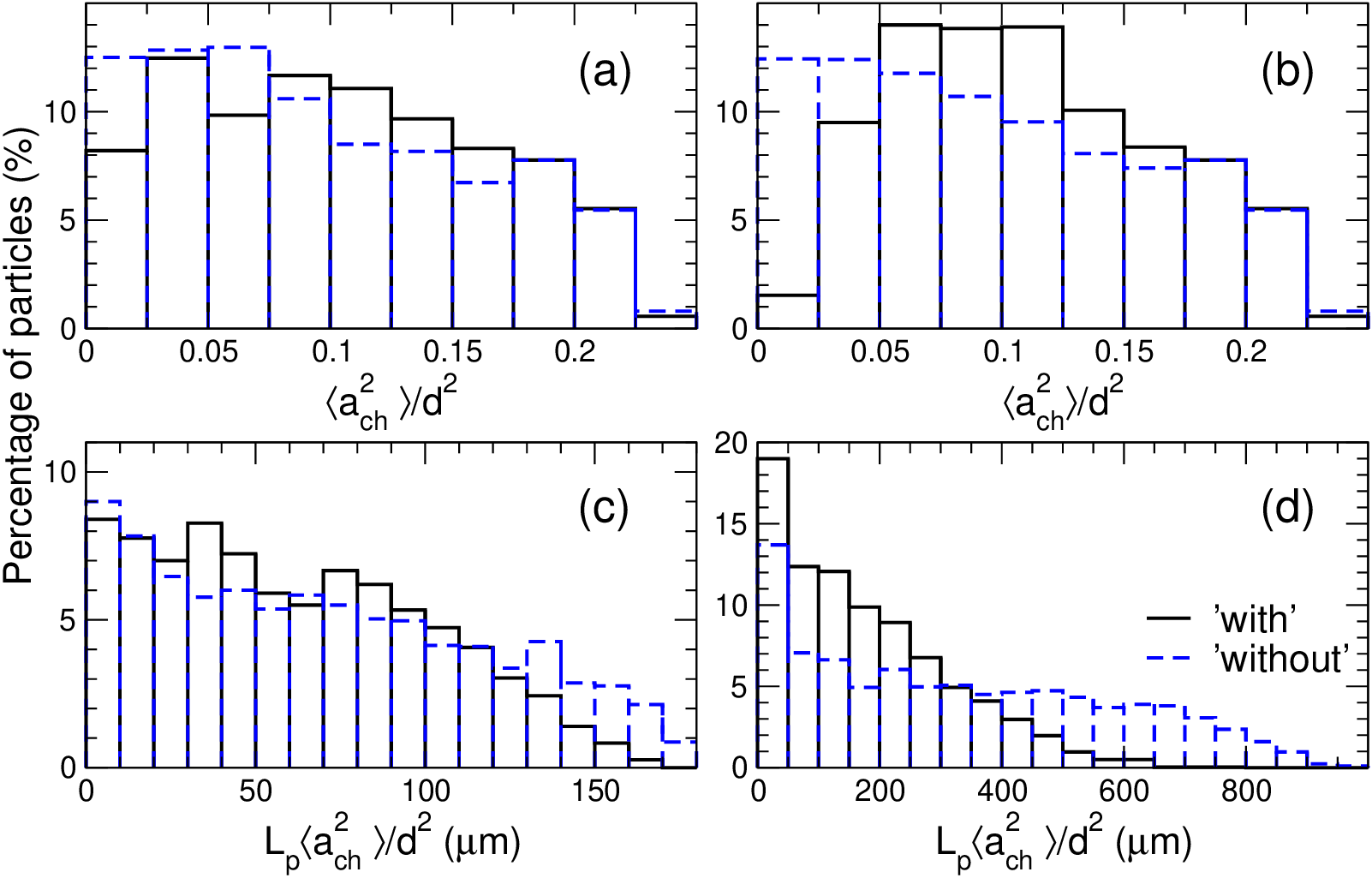}
\caption{
Statistical analysis performed on the initial channelling segments of
the 10 GeV positron trajectories, which were calculated both with
(black solid lines) and without (blue dashed lines) accounting for
ionising collisions in a 1 mm and and 6 mm thick
diamond(110) crystal (left and right columns, respectively).
The vertical axis, common to all graphs, shows the percentage of accepted
particles.
The distributions of particles are shown with respect to:
\textbf{(a)} and  \textbf{ (b):}
average square of the amplitude of channelling oscillations
 (scaled by the square of the (110) interplanar distance $d=1.26$ \AA);
\textbf{c)} and  \textbf{(d):}
product $L_{\rm p}\left\langle a_{\rm ch}^2\right\rangle/d^2 $.
}
\label{Figure07.fig}
\end{figure}
\begin{figure} [h]
\centering
\includegraphics[clip,width=0.9\textwidth]{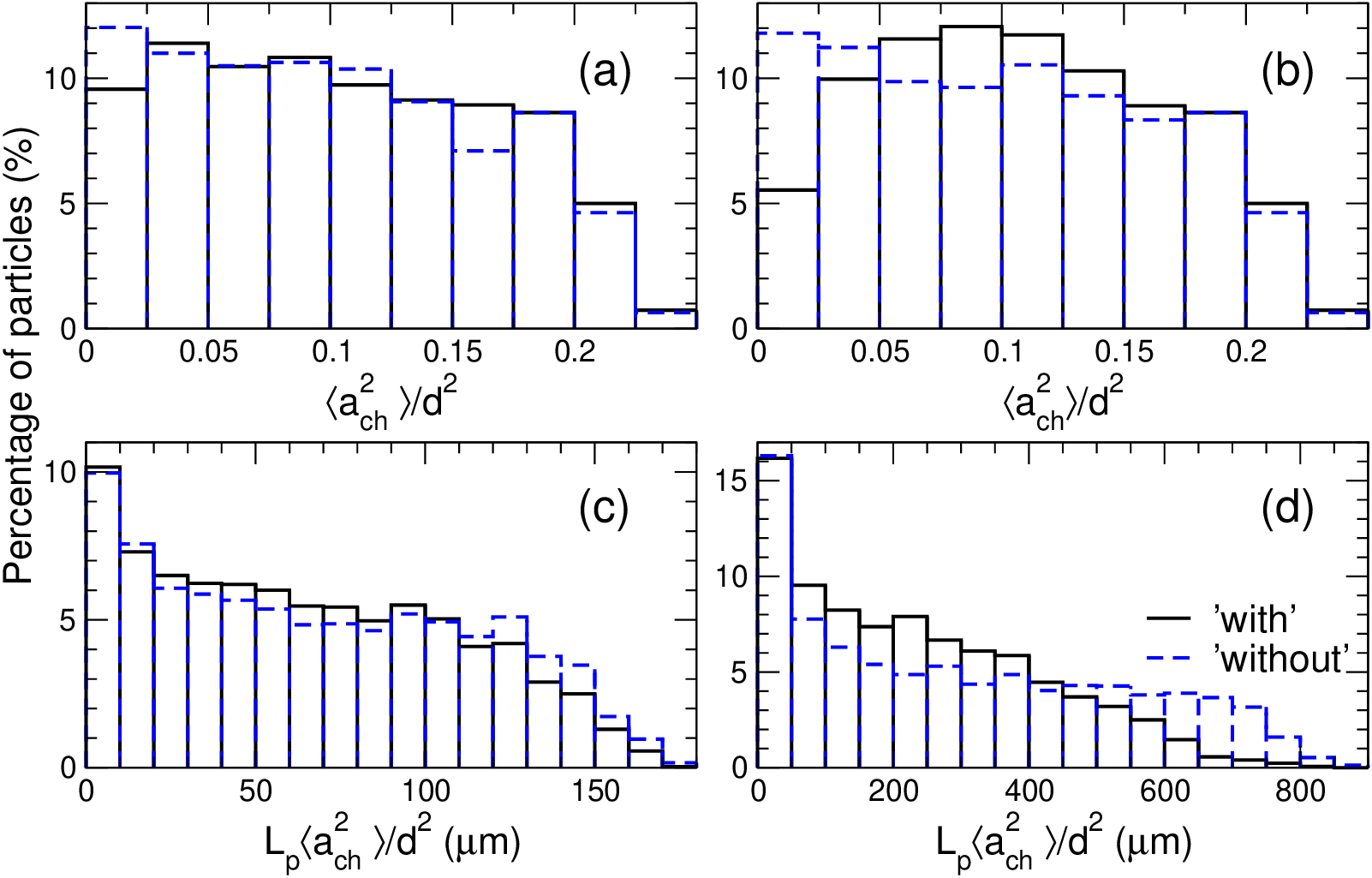}
\caption{
Same as in Fig. \ref{Figure07.fig} but for silicon(110) crystal.
The interplanar distance is 1.92 \AA.
}
\label{Figure08.fig}
\end{figure}

The positron trajectories, simulated with and without accounting for
ionising collisions, were used to statistically analyse
the quantities $\left\langle a_{\rm ch}^2\right\rangle$
and $L_{\rm p}\left\langle a_{\rm ch}^2\right\rangle$.
The results of the exemplary analysis carried out for the (110) channels
are presented as histograms in Figs.
\ref{Figure07.fig} and
\ref{Figure08.fig} for the diamond and
silicon crystals, respectively.
In both figures, the graphs on the left and right refer to crystal
thicknesses of 1 and 6 millimeters, respectively.
The vertical axis is common to all graphs and represents the fraction of
accepted particles, measured in percentages, with respect to the
aforementioned quantities.
The solid and dashed line bars of the histogram correspond to the data
calculated with and without account for the ionising collisions.

The percentage of accepted particles versus the average square of
the amplitude of channelling oscillations is shown in graphs (a) and (b).
The values of $\left\langle a_{\rm ch}^2\right\rangle$ are scaled
by $d^2$ ($d$ is the value of the (110) interplanar spacing,
see the captions).
The histogram stops at 0.25 since $\left(a_{\rm ch}\right)_{\max}=d/2$.
It is seen that the redistribution of particles due to
collisions is most evident for the lighter, thicker crystal,
Fig. \ref{Figure07.fig}(b): the number of particles with small
amplitudes ($\left\langle a_{\rm ch}^2\right\rangle/d^2 \leq 0.05$)
decreases, while the number in the range 0.05–0.175 increases.

Graphs (c) and (d) in the figures illustrate the impact
on the estimated radiation intensity
due to the competition between two opposing tendencies:
the decrease in $L_{\rm p}$ and the increase in
$\left\langle a_{\rm ch}^2\right\rangle$.
When the ionising collisions are accounted for, particles are
redistributed between bins; those with larger
$L_{\rm p}\left\langle a_{\rm ch}^2\right\rangle$
values migrate to bins with lower values.
The extent to which this affects the intensity can be characterised
by the ratio $\xi_{\rm est}$ of the sums on the right-hand side of
Eq. (\ref{CaseStudy_II:eq.01}),
calculated for the 'without' and 'with' datasets.
For diamond targets with thicknesses of 1 and 6 mm, the values of
this ratio are 1.14 and 1.80, respectively.
The corresponding values for the silicon crystal are 1.05 and 1.24,
respectively.
These values of $\xi_{\rm est}$ correlate with the ratios $\xi_{150}$
listed in Table \ref{Table.01}.

The arguments presented above can be used to estimate the intensity of
radiation emitted within the large cone $\theta_0=150\, \mbox{$\mu$rad}
\approx 3/\gamma$, which collects practically all radiation emitted by
accepted particles.
In this case, the radiation intensity scales linearly with the number of
channelling periods $N_{\rm ch}$, which is proportional to the
penetration length $L_{\rm p}$.
However, for a small emission cone $\theta_0$ this scaling should be
modified.
Indeed, under the assumption that channelling oscillations are harmonic,
the peak radiation intensity scales as $N_{\rm ch}^2\propto L_{\rm p}^2$
as $\theta_0 \to 0$.
This factor is due to the constructive interference of electromagnetic
waves emitted from each period.
Then, for a small enough cone with an angle
$\theta_0 = 25\, \mbox{$\mu$rad} \approx 0.5/\gamma$
the scaling factor $N_{\rm ch}^s$ with $1 <s <2$
can be expected.
In this case, the factor $L_{\rm p}$ on the right-hand side of Eq.
(\ref{CaseStudy_II:eq.01}) must be substituted with $L_{\rm p}^s$,
which increases the contribution of trajectories with greater
penetration distances to the sum.
In particular, the increase in radiation intensity emitted by particles
that channel through the entire crystal is greater when ionisation
collisions are disregarded, due to the larger number of such particles
(see the rightmost bins in graphs a) and c) of Figs.
\ref{Figure07.fig} and  \ref{Figure08.fig}).
Consequently, the ratios of the peak intensities calculated
for $\theta_0 = 25$ $\mu$rad are close to the ratios of the number
of channelling particles calculated at the crystal exit
(compare the values of $\xi_{25}$ and $\xi_{\rm ch, 0}$ in
Table \ref{Table.01}).

\section{Conclusions  \label{Conclusions}}

A quantitative analysis has been reported on the impact of inelastic collisions
with crystal atoms on the channeling efficiency and radiation emission of
10 GeV electron and positron beams incident on diamond and silicon single
crystals oriented along the (110) and (111) planar directions.
The crystal thickness along the incident beam was up to 1 mm for electrons
and 6 mm for positrons.

The simulations were performed using the \MBNExplorer software package
within the framework of classical relativistic molecular dynamics.
Events of inelastic scattering were accounted for according to their
probabilities.

The role of the ionising collisions was elucidated through calculations
of the channeling fractions versus penetration distance and the spectral
distribution of the channeling radiation, with and without account for
these collisions.
Generally, the ionising collisions change these dependencies enhancing
the dechanneling rate and decreasing the radiation intensity.

To elucidate the role of ionising collisions, calculations were
performed for channeling fractions versus penetration distance, as well
as for the spectral distribution of emitted radiation, both with and
without accounting for these collisions.
Generally, ionising collisions alter these dependencies, enhancing the
dechanneling rate and decreasing the radiation intensity.

The impact of ionising collisions on electrons is relatively small,
at less than ten percent for diamond targets and a few percent for heavier
silicon crystals.
These comparatively small values are due to the fact that the channeling
motion of electrons occurs in the vicinity of atomic planes.
In this regime, the change in the particle's transverse energy is mainly
due to elastic scattering in the static field of the atoms, rather than
inelastic scattering from atomic electrons.
This feature has a direct practical consequence: numerical simulations
carried out within the framework of classical relativistic molecular
dynamics without accounting for ionising collisions produce a similar
result, but are much faster.

For positrons, the alteration in the dependencies is much more pronounced.
They channel between atomic planes and consequently experience more
distant collisions, in which inelastic scattering dominates.
The decrease in the dechanneling rate is much larger for a diamond crystal
than for a silicon crystal, due to the higher volume density of electrons
in the former.
For example, the number of positrons leaving a 6-millimetre-thick diamond
(110) crystal in  the channeling mode decreases by a factor of five when
accounting for ionisation,
whereas for a silicon (110) crystal of the same thickness, this factor is
approximately two.
Nevertheless, an increase in the dechanneling rate does not directly
result in a comparable decrease in the peak intensity of the channeling
radiation.
This is the case for radiation emitted within a narrow cone along the
incident beam, with an opening angle of much less than the natural
emission cone of $\gamma^{-1}$.
For a wide cone, where the opening angle is much greater than the natural
emission cone, the decrease in intensity due to ionising collisions is
notably smaller.
A detailed quantitative analysis of these features is presented in the
paper.

\section*{Acknowledgements}

The work was supported by the European Commission through
the Horizon  Europe EIC-Pathfinder-Project TECHNO-CLS
(Project No. 101046458).
We acknowledge the Frankfurt Center for Scientific Computing (CSC) for
providing computer facilities.

\vspace*{0.3cm}
\noindent
\textbf{Competing Interests.}
The authors do not declare any conflicts of interest, and there is no
financial interest to report.

\vspace*{0.3cm}
\noindent
\textbf{Author Contribution.}
AVK and AVS contributed equally to the paper.

\appendix

\section{Statistical analysis of the electron trajectories
\label{Appendix01}}

In this appendix, we discuss the physical origin of the difference
in the spectral distribution profiles of the channeling radiation
emitted in the (110) and (111) channels, see Figs.
\ref{Figure02.fig} and \ref{Figure03.fig}.
This is achieved through a statistical analysis of simulated
trajectories, enabling the average parameters of the channeling
oscillations (specifically, the period and the square of the transverse
velocity) to be related to the characteristic energies of the emitted
photons.
The results of the analysis are similar for diamond and silicon crystals;
therefore, we present the diamond case study only.

Prior to this, we briefly summarise the main features of undulator
radiation from a planar undulator
(see reviews
\cite{AlferovBashmakovCherenkov1989,Barbini_EtAl_1990} for more details).

In an ideal planar undulator, the ultra-relativistic charged particle
moves along the harmonic trajectory $y(z) = a\cos(2\pi z/\lambda + \phi)$
with constant amplitude $a$ and period $\lambda$.
The electromagnetic waves emitted by the particle within each period
interfere constructively at certain frequencies
$\om_n$ ($n=1,2,\dots$), which are integer multiples of
the fundamental (first) harmonic frequency $\om_1$.
Consequently, for each value of the emission angle $\theta$ the
spectral distribution consists of a set of equally spaced peaks.
The photon energy corresponding to the fundamental harmonic
can be calculated as follows (see, e.g.,
\cite{Elleaume:RevSciInstrum_v63_p321_1991,CLS-book_2022}):
\begin{eqnarray}
\hbar\om_1
\approx
{ 9.5 \E^2  \over \lambda}
{1 \over 1 + K^2/2 + (\gamma\theta)^2 }\,.
\label{Appendix01:eq.01}
\end{eqnarray}
The factor 9.5 originates from unit conversion:
the particle's energy $\E$ is in GeV, $\lambda$ in $\mu$m,
and the value of $\hbar\om_1$ is in MeV.
The undulator parameter $K$ is related to the particle's average
transverse velocity squared:
\begin{eqnarray}
K^2 = 2\gamma^2 {\langle v_{\perp}^2 \rangle \over c^2}
\label{Appendix01:eq.02}
\end{eqnarray}
where $\gamma$ is the relativistic  Lorentz factor and $c$ is
the speed of light.
The value of $K$ determines the number of emitted harmonics:
for $K^2 \ll 1$, the radiation is predominantly emitted in the
first harmonic; for $K>1$, the number $n$ of harmonics is
estimated as $n \sim K^3$.

In the case of forward emission with an angle of zero, radiation occurs
only in the odd harmonics.
For small emission cones, $(\theta_0\gamma)^2 \ll 1$, the peaks in the
spectral distribution, integrated over $\theta$, are still well
separated and
the peak values of the odd harmonics are higher than those of the even
harmonics.
At $(\gamma\theta_0)^2 > 1$, the peaks are broadened enough to start
overlapping and become less distinguishable.

For positrons, channeling oscillations are nearly harmonic; therefore,
the radiation emission spectra exhibit the aforementioned undulator
radiation features
(see Figs. \ref{Figure05.fig} and \ref{Figure06.fig}).

\begin{figure} [h]
\centering
\includegraphics[clip,width=0.8\textwidth]{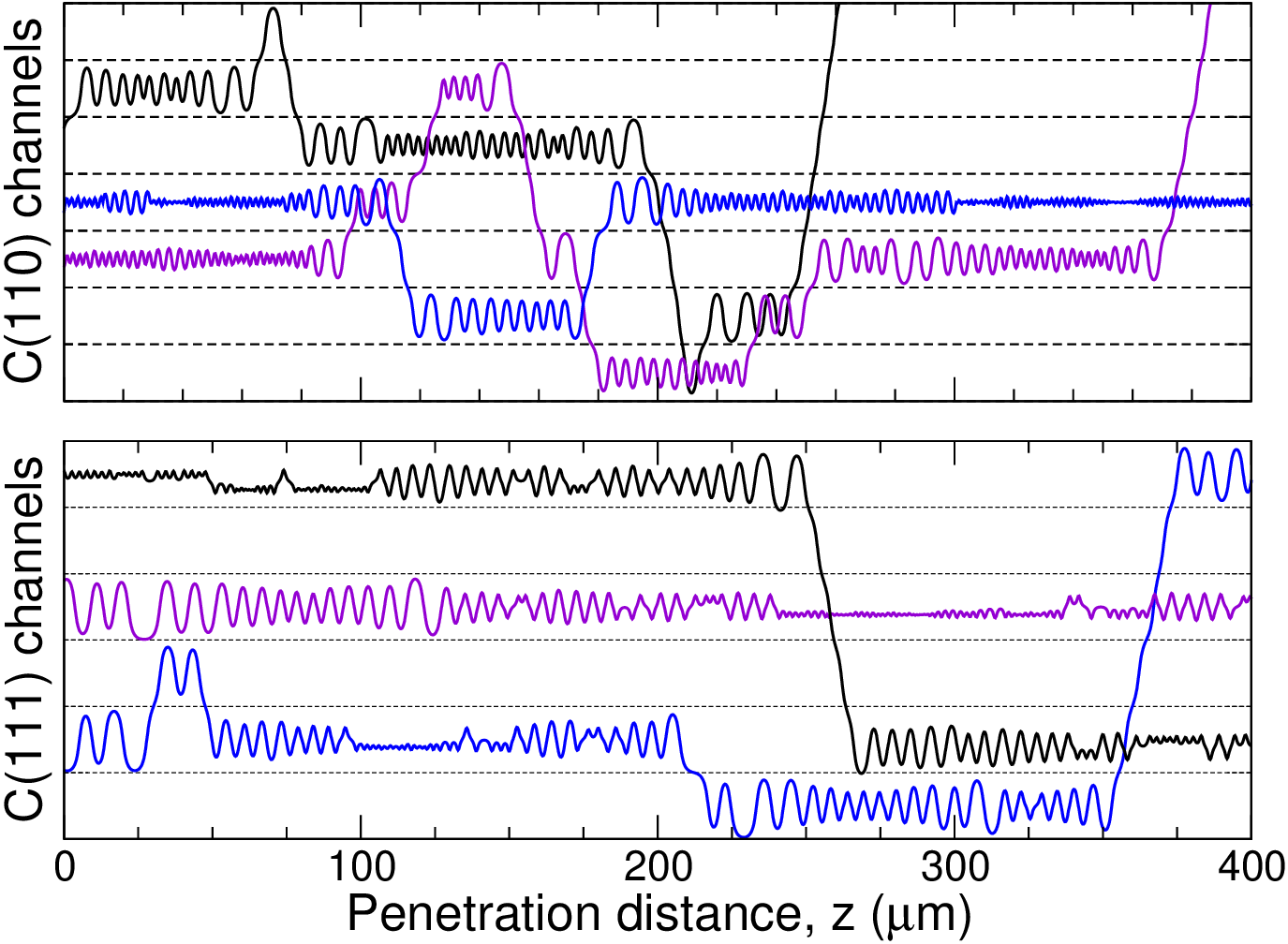}
\caption{
Selected electron trajectories in the (110) (\textbf{top} graph)
and (111) (\textbf{bottom} graph) channels in diamond.
Dashed lines visualise the channel boundaries.
The channeling segments illustrate the anharmonic character of the
channeling oscillations.
}
\label{Figure09.fig}
\end{figure}

The channeling oscillations of electrons are strongly anharmonic,
resulting in a noticeable change in the profile of channeling radiation.
Figure \ref{Figure09.fig} shows several selected trajectories
of 10 GeV electrons simulated in (110)- and (111)-oriented
diamond crystals (the top and bottom graphs, respectively).
The channeling segments shown in the figure illustrate the variation
in period $\lambda_{\rm ch}$ and amplitude of the oscillations
along the trajectory.
Anharmonicity also leads to the variation in the undulator parameter
$K_{\rm ch}$, as defined in Eq. (\ref{Appendix01:eq.01}).

The statistical analysis of these variations and their
impact on the fundamental harmonic, Eq. (\ref{Appendix01:eq.01}),
was conducted as follows.
For each channel, a set of approximately 3000 trajectories was
randomly selected.
The channeling segments were identified in each trajectory, and the
average values of $\langle \lambda_{\rm ch} \rangle$ and
$\langle K_{\rm ch}^2 \rangle$ were calculated along each segment.
These values were used in (\ref{Appendix01:eq.01}) to
characterise the energy $\hbar\om_{\rm ch}$
of the first harmonic of channeling radiation
emitted from the segment in the forward direction (i.e. for $\theta=0$).

\begin{figure} [h]
\centering
\includegraphics[clip,width=0.8\textwidth]{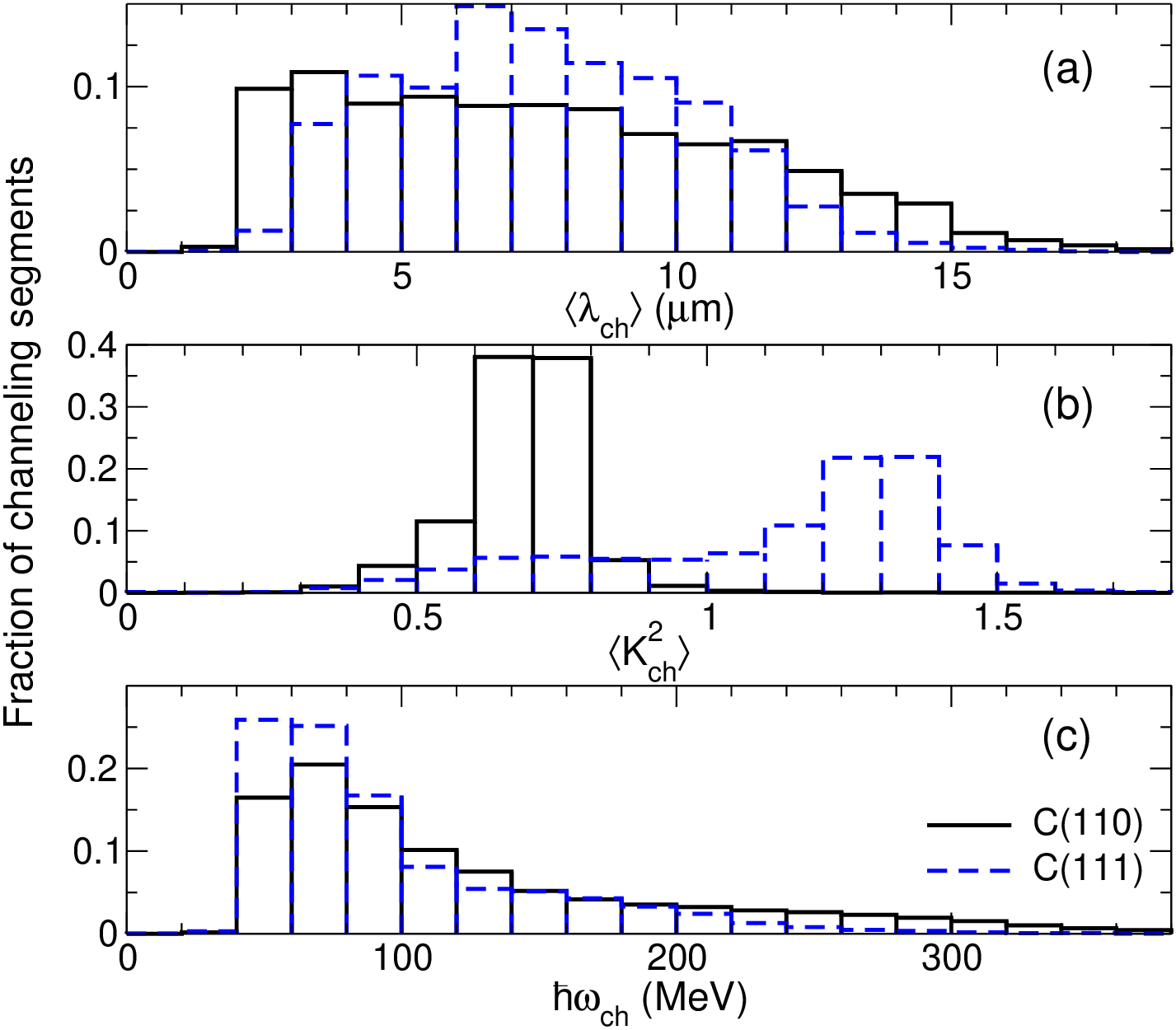}
\caption{
The distributions of the channeling segments
with respect to average values of
\textbf{(a)} the channeling period $\langle \lambda_{\rm ch} \rangle$,
\textbf{(b)} the undulator parameter squared
$\langle K_{\rm ch}^2 \rangle$,
and \textbf{(c)} the energy of the first harmonic $\hbar\om_{\rm ch}$
for $\theta=0$,
calculated for 10 GeV electrons in (110)- and (111)-oriented
diamonds, solid and dashed lines, respectively.
}
\label{Figure10.fig}
\end{figure}

The results of these calculations are presented in
Fig. \ref{Figure10.fig} in the form of
histograms for the fractions of the channeling segments
(approx. $2\times 10^4$ segments were analysed for each channel).
Graphs (a)-(c) show the distribution of the fractions with respect to
$\langle \lambda_{\rm ch} \rangle$, $\langle K_{\rm ch}^2 \rangle$ and
$\hbar\om_{\rm ch}$, respectively.

For both channels, the value of $\langle K_{\rm ch}^2 \rangle/2$ is less
than one, so it is primarily the variation in
$\langle \lambda_{\rm ch} \rangle$
that affects the value of $\hbar\om_{\rm ch}$.
Comparing the distributions for the (110) (solid lines)
and (111) (dashed lines) channels, it can be seen that a more even
distribution of  $\langle \lambda_{\rm ch} \rangle$ in the (110) channel
leads to a broader range of values for the $\hbar\om_{\rm ch}$ values.
Figure \ref{Figure10.fig}(b) shows that, for all segments,
the value of $\langle K_{\rm ch}^2 \rangle$ is well below one for the
(110) channel.
However, for a large fraction of the channeling segments, this value is
greater than one for the (111) channel.
This difference in the undulator parameter explains why the signals of
several first harmonics are seen in the spectral distribution of radiation
emitted in the (111) channel.

\section*{References}

\end{document}